\documentclass[aps,prl,amssymb,groupedaddress,nofootinbib]{revtex4}

\usepackage[english]{babel}
\usepackage{amsfonts}
\usepackage{amssymb}
\usepackage{epsfig}
\usepackage{amsmath}
\usepackage{fancyhdr}
\usepackage{textcomp}
\usepackage{setspace}

\begin{document}

\title{Cosmic String Power Spectrum, Bispectrum and Trispectrum}

\author{D.M.~Regan}

\author{E.P.S.~Shellard}

\affiliation{Centre for Theoretical Cosmology,\\
Department of Applied Mathematics and Theoretical Physics,\\
University of Cambridge,
Wilberforce Road, Cambridge CB3 0WA, United Kingdom}

\date{\today}

\begin{abstract}
We use analytic calculations of the post-recombination gravitational effects of cosmic strings to estimate
the resulting CMB power spectrum, bispectrum and trispectrum.   We place a particular emphasis on
multipole regimes relevant for forthcoming CMB experiments,
notably the Planck satellite.    These calculations use a flat sky approximation, generalising
previous work by integrating string contributions from last scattering to the present day, finding
the dominant contributions to the correlators for multipoles $l> 50$.   We find a well-behaved shape
for the string bispectrum (without divergences) which is easily distinguishable from 
inflationary bispectra which possess significant acoustic peaks.   We estimate that the
nonlinearity parameter characterising the bispectrum is approximately $f_{NL} \sim -20$ (given
present string constraints from the CMB power spectrum.
We also apply these unequal time correlator methods to calculate the trispectrum for
parrallelogram configurations, again valid over a large range of angular scales relevant
for WMAP and Planck, as well as on very small angular scales.   We find that, unlike the
bispectrum which is suppressed by symmetry considerations, the trispectrum for cosmic
strings is large.    Our current estimate
for the trispectrum parameter is $\tau_{NL} \sim 10^4$, which may provide one of the strongest
constraints on the string model as observational estimates for the trispectrum improve.
\end{abstract}

\pacs{1}

\maketitle
\section{Introduction}

Cosmic strings are a common feature in fundamental cosmology scenarios, such as brane inflation \cite{Tye_or_review}, and
they appear to be generic in realistic grand unified theories \cite{Jeannerot}.  Cosmic strings
leave a distinct `line-like' signature in the cosmic microwave background (CMB) \cite{Gott,KAISERSTEBBINS} which
offers one of the best prospects for their detection.   Strings create this imprint after recombination by perturbing photons through relativistic gravitational effects.  While inflationary fluctuations are believed
to dominate the overall CMB power spectrum, current constraints suggest that up to 10\% of the
signal could be contributed by cosmic strings \cite{Battye, Kunzetal}. A higher proportion
is incompatible with WMAP because the cosmic string power spectrum around
multipoles $l \approx 200$ is dominated by metric fluctuations with a relatively featureless spectrum, rather than the pre-recombination acoustic peaks characterising inflation.  The gravitational strength of cosmic strings
is determined by the parameter $G\mu = (\eta/m_{Pl})^2$ which is the ratio of the string tension $\mu$ to
the square of the Planck mass $m_{Pl}$.     The present WMAP limit on the string contribution to
the CMB translates into a strong constraint on the parameter $G\mu \leq 2.5 \times 10^{-7}$
 \cite{Battye}.
\par
Inflationary CMB fluctuations in the standard picture are very nearly Gaussian, so higher order correlators offer the prospect of  further differentiation between these and competing signals from
cosmic strings.
The spherical harmonic transform of the three-point CMB correlator is the bispectrum $B_{l_1 l_2 l_3}$.   Usually discussions of the bispectrum are simplified further by focusing on the nonlinearity parameter $f_{NL}$ which is roughly the ratio of the three point correlator to the square of the two-point correlator,
that is, $f_{NL}\sim \langle \zeta\zeta\zeta \rangle /\langle\zeta\zeta\rangle^2$ where the $\zeta$ are the primordial curvature fluctuations (dimensionless) which seed the CMB anisotropies. For a perturbative theory like inflation, we expect the second-order fluctuations $\zeta^{(2)}$ to be constructed from convolutions of the linear perturbations $\zeta^{(2)} \sim \zeta^{(1)}* \zeta^{(1)}$.  This implies that
the leading order term in the three-point correlator can be expected to behave as
$\langle \zeta^{(1)}\zeta^{(1)}\zeta^{(2)}\rangle \sim \langle \zeta^{(1)}\zeta^{(1)} \rangle^2$, that is, $f_{NL} \sim 1$.   In fact, for standard single field inflation there is considerable further suppression from slow-roll and the primordial signal $f_{NL}\approx {\cal O}(0.01)$ \cite{Maldacena}.  One can only obtain a significantly larger $f_{NL}$ in more exotic inflationary models with multiple scalar fields or non-canonical kinetic terms (see, for example, the reviews in \cite{Chen, Fergusson}).  Similar arguments
apply to the trispectrum or four-point correlator characterised by $\tau_{NL} \sim \langle \zeta\zeta\zeta\zeta\rangle/\langle\zeta\zeta\rangle^3$ with $\tau_{NL} \lesssim  1$ for standard single field inflation.
\par
In contrast, the cosmic string signature has an inherently non-perturbative origin so we do not expect
$f_{NL}$ to be small nor, more particularly, $\tau_{NL}$.    For the higher order correlators from cosmic strings (measured relative
to the string power spectrum $\langle \zeta\zeta\rangle_{cs}$), the relevant scaling with the parameter $G\mu$ is $f_{NL} \sim {\cal A} (G\mu)^{-1}$ \cite{Hindmarsh2009} and, as we shall discuss here, $\tau_{NL} \sim {\cal B}(G\mu)^2$
 with coefficients ${\cal A}$ and ${\cal B}$ determined by geometric and dynamical considerations.
There is a suppressed amplitude ${\cal A}$ for the bispectrum (due to symmetry
cancellations \cite{Hindmarsh2009}), and we contrast this here with one of the key results
of this paper which is the much larger relative value of
${\cal B}$ for the trispectrum.  Here, we shall compare the higher order correlators from strings we
calculate with  the dominant inflationary power spectrum $\langle \zeta\zeta\rangle_{inf}$ and we will show
that the trispectrum with $\tau_{NL}\ggg 1$ is  the more likely correlator to produce the strongest
 constraints on cosmic strings (or with which to identify them).
\par
In this paper, we use analytic calculations of the post-recombination string signature to estimate
the power spectrum, bispectrum and trispectrum relevant for forthcoming CMB experiments,
particularly the Planck satellite.    We generalise previous analytic calculations of Hindmarsh and
collaborators for the power spectrum \cite{Hindmarsh} and bispectrum \cite {Hindmarsh2009} based on a flat sky approximation.  While this
was an important first step, the summation employed was essentially confined only to strings close
to the surface of last scattering and it is only relevant for very small angular scales.   Here,
we note that the dominant contribution on large angular scales (as well as very small angles)
comes from the same gravitational GKS effects but from strings at late times well after last scattering.
By integrating in time over the unequal time correlators, we obtain an approximate bispectrum
valid for all $l \gtrsim 50$ (where the flat sky approximation breaks down), i.e. useful
for both WMAP and Planck.   We also apply these
unequal time correlator methods to the first calculation of the trispectrum (for parralelogram
configurations), which is again valid over a large range of angular scales.   Our bispectrum calculation
also differs from ref.~\cite{Hindmarsh2009} by eliminating the divergences they find in their Gaussian integrals for flattened triangles.  We note that these are actually cut-off by the behaviour of other terms in the integrand (due to momentum constraints), yielding a finite result in this regime.   The cross-sectional
shape of the bispectrum we present is relatively
featureless, except for a finite rise near the edges (for flattened triangles) and
suppression towards the corners because of causality contraints (squeezed triangle limit).  This
well-behaved shape is suitable for the bispectrum estimation methods being developed for Planck
and is easily distinguishable  from bispectra predicted
by inflation because of the absence of acoustic peaks \cite{Fergusson}. Of course, these analytic calculations neglect important recombination
effects which also provide significant contributions to the string bispectrum for $500\lesssim l
\lesssim 2000$.
However, determining the extent to which these contributions can confuse the
non-Gaussian `line-like' signature of cosmic strings will be the subject of future study \cite{FergLandetal}.
\par
These results for the string bispectrum and trispectrum are important for CMB experiments and should be
valid and dominant at both
large and very small angular scales.    While the Planck satellite does not have the resolution
to see individual string signatures, it should be possible to obtain statistically significant
constraints on cosmic strings that compete with limits on $G\mu$ from the power spectrum,
especially for the trispectrum when these techniques are fully developed. On very small
angular scales the string power spectrum begins to dominate over the inflationary signature
($l>3000$) because it is not influenced by exponentially decaying transfer functions.  Here
direct detection of `line-like' signatures may prove possible provided that experiments can
achieve $\mu K$ sensitivities, as anticipated by AMI and ACT, for example. Again, searching
in larger data sets for higher order correlators may provide statistically more significant
results.

\section{Gott-Kaiser-Stebbins effect}
In this section we calculate the Gott-Kaiser-Stebbins (GKS) effect which is expected to give the principle contribution to cosmic strings on subhorizon scales where we may use the flat sky approximation and can ignore the response of cosmological fluids. With these approximations the integrated Sachs Wolfe effect is equal to the temperature discontinuity in the CMB. In fact we shall show in this paper that it is the `late time small angular effect' induced by the GKS effect that also dominates large angular scales for cosmic strings. The discussion here otherwise follows that of \cite{Hindmarsh2009}.
\par
 Consider a photon with $4$-momentum $p_{\mu}=(E,0,0,E)$ and a string of coordinates $X^{\mu}(\sigma,t)$, where $(\sigma, t)$ are the worldsheet coordinates of the string (with the gauge chosen such that $t$ corresponds to the (conformal) time coordinate). The unperturbed geodesics can be written as
\begin{eqnarray*}
Z^{\mu}=x^{\mu}+\lambda p^{\mu}
\end{eqnarray*}
The perturbation to the energy along the photon path (which gives the ISW effect) is
\begin{eqnarray}
\delta p_{\mu}=-\frac{1}{2}\int_{\lambda_1}^{\lambda_0}h_{\nu\rho,\mu}(Z(\lambda))p^{\nu}p^{\rho} d\lambda
\end{eqnarray}
where $h_{\mu\nu}$ is the metric perturbation which in general must be calculated. However, in
refs.~\cite{Stebbins,Hindmarsh} it was pointed out that we can simplify the discussion by considering
\begin{eqnarray*}
\mathbf{\nabla}_{\perp}^2 \delta p_{\mu}
\end{eqnarray*}
where $\mathbf{\nabla}_{\perp}$ represents partial derivatives with respect to the transverse coordinates to the unperturbed geodesics of the photons.
\newline Now suppose that the photon motion is in the $z$-direction. This implies that $(\partial_t-\partial_z)f(Z)=E^{-1}\frac{df}{d\lambda} $. Also we note the following
\begin{eqnarray*}
\mathbf{\nabla}^2 \delta p_{\mu}&=&(\partial_t^2-\partial_z^2-\partial^2)\delta p_{\mu}\\
&=&-\frac{1}{2}\int_{\lambda_0}^{\lambda_1}d\lambda \left[      (\partial_t+\partial_z)(\partial_t-\partial_z)     -\partial^2\right]h_{\nu\rho,\mu}(Z(\lambda))p^{\nu}p^{\rho}
\end{eqnarray*}
Now suppose we use the harmonic gauge for the metric perturbations. Then we have $\partial^2 h_{\mu \nu}=16\pi G(T_{\mu\nu}-1/2 \eta_{\mu\nu}T)$ which implies that we have
\begin{eqnarray*}
\mathbf{\nabla}^2 \delta p_{\mu}&=&\left[       \frac{1}{2E}(\partial_t+\partial_z)h_{\nu\rho,\mu}p^{\nu}p^{\rho}       \right]^{\lambda_0}_{\lambda_1}+8\pi G\partial_{\mu}\int_{\lambda_0}^{\lambda_1}d\lambda  T_{\nu\rho}p^{\nu}p^{\rho} \,.
\end{eqnarray*}
Next we define $\hat{p}^{\mu}=p^{\mu}/E$. Since we are interested in the energy we focus on the $\mu=0$ component. We then obtain\footnote{Note that the formula  $(\partial_t-\partial_z)f(Z)=E^{-1}\frac{df}{d\lambda} $ generalises for a general photon path to $(\partial_t+\hat{p}^i\partial_i)=E^{-1}\frac{d}{d\lambda} $ }
\begin{eqnarray*}
T_{\rho i,0}\hat{p}^i&=&\frac{1}{E}\frac{dT_{\rho i,0}\hat{p}^i}{d\lambda} -T_{\rho i,j}\hat{p}^i\hat{p}^j\\
\implies T_{\rho\nu,0}\hat{p}^{\nu}&=&\frac{1}{E}\frac{dT_{\rho i,0}\hat{p}^i}{d\lambda} -T_{\rho i,j}\hat{p}^i\hat{p}^j +T_{\rho j,j}\\
&=&\nabla_{\perp}^i T_{\rho i}+\frac{1}{E}\frac{dT_{\rho i,0}\hat{p}^i}{d\lambda}\,,
\end{eqnarray*}
where we have used  $T_{\rho j,j}=T_{\rho 0,0}$ in the second line and we define $\nabla_{\perp}^i =\partial^i-\hat{p}^i\hat{p}^j\partial_j$.
This leads to
\begin{eqnarray*}
\mathbf{\nabla}^2 \frac{\delta E}{E}=8\pi G\int d\lambda E \nabla_{\perp}^i T_{i \rho}\hat{p}^{\rho}+\frac{1}{2}\left[    (\partial_t+\partial_z)\partial_t\hat{h}-16\pi GT_{\rho ij}\hat{p}^i\hat{p}^{\rho}\right]_{\lambda_0}^{\lambda_1}\,.
\end{eqnarray*}
The terms in square brackets are boundary terms. These may be important at decoupling however due to the finite thickness of the last scattering surface we expect these fluctuations to be smeared out on the scales of interest and so we neglect these terms. For a string source\footnote{The energy momentum tensor in the conformal gauge is $T^{\mu\nu}=\int dt d\sigma( \dot{X}^{\mu}\dot{X}^{\nu}-{X'}^{\mu}{X'}^{\mu})\delta^{(4)}(x-X)$} in the light cone gauge ($X^0+X^3=t$) this yields
\begin{eqnarray*}
\mathbf{\nabla}_{\perp}^2\delta=-8\pi G \mu \int d\sigma \dot{X}.\mathbf{\nabla}_{\perp}\delta^{(2)}(\mathbf{x-X})
\end{eqnarray*}
where the quantities are evaluated at $t_r=t+z-X^3(\sigma,t_r)$. The wavenumber $\mathbf{k}$ is related to the multipole moment, $l$, for $l\gtrsim 60$ via $k^2\approx l(l+1)/(t_0-t_r)^2\approx l(l+1)/t_0^2$, where $t_0$ is the conformal time today.
\newline
In Fourier space we have
\begin{eqnarray}\label{stringsource}
-k^2 \delta_k(t_r)=i8\pi G \mu k^A \int d\sigma \dot{X}^A(\sigma,t_r) e^{i\mathbf{k.X}(\sigma,t_r)}
\end{eqnarray}
where $A=1,2$ runs over the transverse coordinates.

\section{Power Spectrum}
\subsection{Power Spectrum on small angular scales $l\gtrsim 500$}
In order to calculate the power spectrum we find the  unequal time correlator for density perturbations formed at different light cone crossing times and then we sum these up between the last scattering surface and today to get the total power spectrum $P(k)$. Defining the unequal time correlator as
 \begin{eqnarray}\label{uetcpower}
 P(k_1,t_1,t_2)=(2\pi)^2\delta^{(2)}(\mathbf{k_1+k_2})\langle\delta_{\mathbf{k_1}}(t_1)\delta_{\mathbf{k_2}}(t_2)\rangle\,,
 \end{eqnarray}
 we can find it by integrating over the string source terms (\ref{stringsource}) at the given times, 
\begin{eqnarray}
P(k,t_1,t_2)=(8\pi G \mu )^2\frac{k^A k^B}{\mathcal{A}k^4}\int d\sigma d\sigma'\left<  \dot{X}^A(\sigma,t_1)\dot{X}^B(\sigma',t_2) e^{i\mathbf{k.(X(\sigma,t_1)-X(\sigma',t_2))}}\right>\,,
\end{eqnarray}
where $\mathcal{A}=(2\pi)^2 \delta(0)$ is a formal area factor.
We obtain the power spectrum through the sum from last scattering $t_{lss}$  to today $t_0$:
\begin{eqnarray*}
P(k)=\sum_{t_{lss}}^{t_0}\sum_{t_{lss}}^{t_0}dt_1 dt_2 P(k,t_1,t_2)
\end{eqnarray*}
\par
Now we use the following three assumptions (for the network at a given time as in ref.~\cite{Hindmarsh2009}): (i)
The string ensemble is a Gaussian process, i.e. we can find all the correlation functions in terms of two point correlators.  (ii)
We have reflection and translation invariance of the transverse coordinates.  (iii)
We have reflection and translation invariance of the worldsheet coordinates.   This means
that for equal time correlators we can write

Therefore we can write (for equal time correlators)
\begin{eqnarray*}
\left<  \dot{X}^A(\sigma,t)\dot{X}^B(\sigma',t)\right>&=&\frac{\delta^{AB}}{2}V(\sigma-\sigma',t)\\
\left< \dot{X}^A(\sigma,t){X'}^B(\sigma',t)\right>&=&   \frac{\delta^{AB}}{2}M_1(\sigma-\sigma',t)\\
\left<{X'}^A(\sigma,t){X'}^B(\sigma',t)\right>&=&\frac{\delta^{AB}}{2}T(\sigma-\sigma',t)
\end{eqnarray*}
and hence we have
\begin{eqnarray*}
\left< (X^A(\sigma,t)-X^A(\sigma',t))^2\right>&=&\int_{\sigma'}^{\sigma} \int_{\sigma'}^{\sigma} d\sigma_1 d\sigma_2 T(\sigma_1-\sigma_2,t)=:\Gamma(\sigma-\sigma',t)\\
\left< (X^A(\sigma,t)-X^A(\sigma',t))\dot{X}^A(\sigma',t)\right>&=&\int_{\sigma'}^{\sigma} d\sigma_1 M_1(\sigma_1-\sigma,t)=: \Pi(\sigma-\sigma',t)\,.
\end{eqnarray*}
This is an extremely powerful relation since we can use the fact that at small angles, i.e. small distances for correlators we have that $\Gamma(\sigma,t)\propto \sigma^2, V(\sigma,t)\propto \sigma^0, \Pi(\sigma,t)\propto \sigma^2$. In particular we write (for small angles)
\begin{eqnarray*}
\Gamma(\sigma,t)\approx \overline{t}^2\sigma^2, V(\sigma,t)\approx \overline{v}^2, \Pi(\sigma,t)\approx \frac{c_0}{2\hat{\xi}}\sigma^2
\end{eqnarray*}
where $\overline{t}^2=\frac{2}{3}-\overline{v}^2$ and $\hat{\xi}=\tilde{\xi} t$ is the correlation length of the network which scales in (conformal) time, i.e. $\tilde{\xi}=\mbox{constant}$.  At time $t\gtrsim t_{lss}$, we have the following relation between the wavenumber $k$ and the multipole $l$
\begin{eqnarray}\label{wavemultipole}
k\hat{\xi}\approx \frac {l}{500}\frac{t_{lss}}{t}\,,
\end{eqnarray}
that is, the correlation length at time $t\gtrsim t_{lss} $ corresponds to a multipole $500 (t_{lss}/t)$ since the correlation length at last scattering corresponds to a multipole of approximately $l\approx 500$. Therefore, for angular scales below $500$ we cannot integrate back to last scattering and instead to $t_{lss}(500/l)>t_{lss}$.
\par
For unequal times, we have on small angular scales
\begin{eqnarray*}
\tilde{\Gamma}(\sigma,\sigma',t_1,t_2)&=&\left< (X^A(\sigma,t_1)-X^A(\sigma',t_2))^2\right>\approx\overline{t}^2 (\sigma-\sigma')^2+\overline{v}^2(t_1-t_2)^2
 \end{eqnarray*}
Hence, on small angular scales,
\begin{eqnarray*}
P(k,t_1,t_2)\propto e^{-k^2 \overline{v}^2(t_1-t_2)^2/4}
\end{eqnarray*}
and so we have that $P(k,t_1,t_2)\approx P(k,t_1,t_1)  $, such that
\begin{eqnarray*}
P(k)\approx \int_{1}^{t_0/t_{lss}} dt_1 P(k,t_1,t_1)
\end{eqnarray*}
where we use the renormalised time $t/t_{lss}$.
\par
\begin{figure}[htp]
\centering 
\includegraphics[width=82mm]{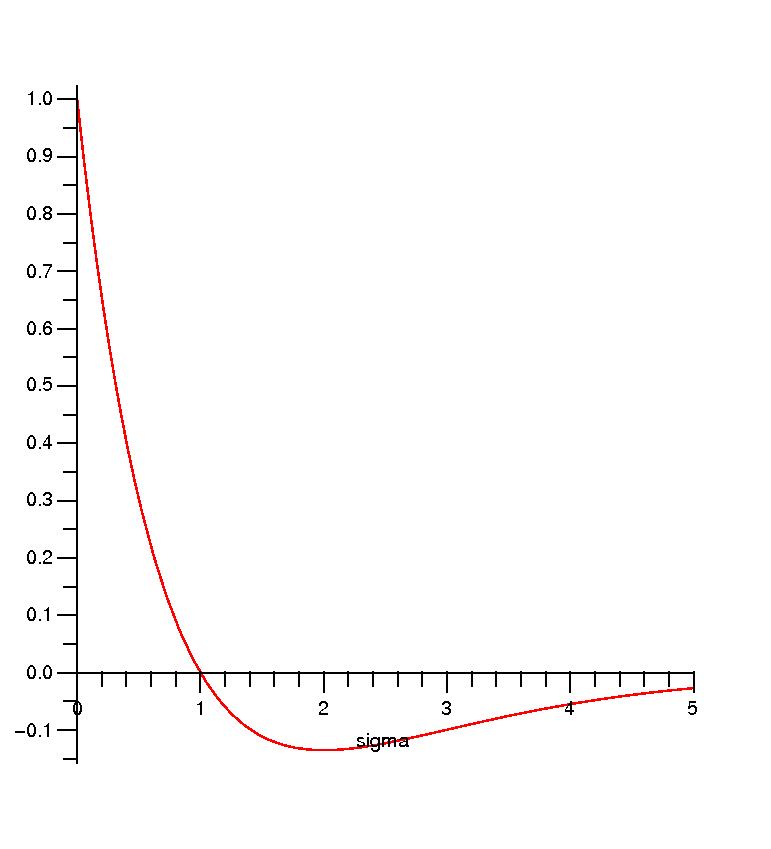}
\caption{Plot of the improved approximation of $V(\sigma,t)/(\overline{v}^2)$. The velocity approaches zero at the correlation length at each time $t$ and displays an anti-correlation which is a result of momentum conservation in the network. The plot is in units of the correlation length.}
\label{fig:velocity}
\end{figure}

Now using
\begin{eqnarray*}
P(k,t,t)=(8\pi G \mu )^2\frac{k^A k^B}{4\mathcal{A}k^4}\int d\sigma_+ \int d\sigma_-\left<  \dot{X}^A(\sigma,t)\dot{X}^B(\sigma',t) e^{i\mathbf{k.(X(\sigma,t)-X(\sigma',t))}}\right>
\end{eqnarray*}
where $\sigma_+=\sigma+\sigma'$ and  $\sigma_-=\sigma-\sigma'$, we approximate the integral as
\begin{eqnarray*}
P(k,t,t)=(8\pi G \mu )^2\frac{\mathcal{L}}{\mathcal{A}}\frac{1}{2 k^2}\int d\sigma_-\left(V(\sigma_-,t)+\frac{k^2}{2}\Pi(\sigma_-,t)  \right)\exp\left(-\frac{k^2}{4}\Gamma(\sigma_-,t)\right)
\end{eqnarray*}
where we note that the range of integration (in $\sigma$) is limited by the length of string in the network. For large wavenumbers, $k\gtrsim 5000$, we may use the small angle approximations as outlined above. However for smaller wavenumbers we can approximate this by constraining the range of $\sigma_-$ to $(-\hat{\xi}/2,\hat{\xi}/2)$, since simulations show that in the range the approximations made above are reasonably accurate for $k\gtrsim 500$. We can make a better approximation motivated by the velocity correlator as found numerically in ref.~\cite{MartShell}\footnote{The velocity correlator here is in the light cone gauge. However the behaviour in any other temporal gauge should follow a similar behaviour.}. In particular we approximate the velocity correlator (see Figure~\ref{fig:velocity})
 to be\footnote{This approximation obeys the constraint $\int d\sigma V(\sigma)=0$ which is a reflection of conservation of momentum in the network.}
\begin{eqnarray*}
V(\sigma,t)\approx \overline{v}^2 \left(1-\frac{|\sigma|}{\hat{\xi}}\right)\exp\left(-|\sigma|/\hat{\xi}\right)
\end{eqnarray*}

Using this approximation we may safely push the range of integration for $V(\sigma,t)$ to $\sigma\in(-\infty,\infty)$. However, using this (better) approximation here makes little difference to the total power spectrum and so we do not use it at this junction. Nevertheless, this approximation becomes very important for the bispectrum and trispectrum.
\par
The second term in the above approximation gives a subdominant contribution to the power spectrum at all angular scales and therefore we discuss only the first term below. The power spectrum is therefore given by
\begin{eqnarray*}
k^2P(k)\approx l^2 C_l\approx &&(8\pi G \mu )^2\frac{\mathcal{L}\hat{\xi}}{\mathcal{A}}\frac{1}{\tilde{\xi}}\int_1^{t_0/t_{lss}} \frac{dt}{t}\int_{0}^{\tilde{\xi} t/2} d\sigma'   \overline{v}^2\frac{\sqrt{\pi}}{2\overline{t}l} \exp\left(-\frac{\overline{t}^2 l^2}{4} {\sigma'}^2\right)\\
\approx &&(8\pi G \mu )^2\frac{\mathcal{L}\hat{\xi}}{\mathcal{A}}\frac{1}{\tilde{\xi}}\frac{\sqrt{\pi}}{2\overline{t}l} \int_1^{t_0/t_{lss}} \frac{dt}{t}\overline{v}^2  \mbox{erf}\left(\frac{\overline{t}l\tilde{\xi} }{4}t\right)
\end{eqnarray*}
since the factor $\mathcal{L}\hat{\xi}/\mathcal{A}\approx \mbox{constant}$ and again we use the renormalised time $t/t_{lss}$ and the renormalised worldsheet coordinate $\sigma'=\sigma/t_{0}$ to simplify notation. For the purposes of finding the total power spectrum we approximate the error function by a constant - in particular, by unity since it will quickly reach this value if it is not already equal to unity at last scattering. Hence, we obtain the small angle power spectrum ($l \gtrsim 500$):
\begin{eqnarray}
k^2 P(k)\approx l^2 C_l\approx (8\pi G \mu )^2\sqrt{\pi}\frac{\mathcal{L}\hat{\xi}}{2 \mathcal{A}}\frac{\overline{v}^2}{\overline{t}}\frac{1}{l\tilde{\xi} }\ln\left( \frac{t_0}{t_{lss}}\right)\sim l^{-1}\,.
\end{eqnarray}

\subsection{Power Spectrum on large angular scales, $l\lesssim 500$}
In ref.~\cite{Hindmarsh}, it was shown that if we compute the large angle power spectrum using the contribution of strings at the last scattering surface then $k^2 P(k)\propto k^2$ or $l^2C_l\propto l^2$
on lengthscales above the correlation length ($l\lesssim 500$). However, we will show in this section that the late time small angle contribution of strings, i.e. the small angle effect of cosmic strings as the network evolves between last scattering and today, is the dominant contribution to the low $l$ part of the spectrum. We find this by integrating the unequal time correlator for all times between $t>t_{lss}$ and $t_0$ and finding where the peak of the spectrum from such times is located. Such contributions are, in effect, superhorizon effects at last scattering which come inside the horizon at later times. Essentially we are tracking the evolution of the peak of the spectrum (located at $l\approx 500$ at last scattering) in time.  Here we simply note that the contributions on superhorizon scales at a time $t=t_H$ fall off
sufficiently rapidly for $k<aH$ that they can be neglected relative to the late time contributions
$ t>t_H$ from subhorizon strings.

Integrating the string sources over all times since last scattering we obtain the power spectrum relevant
on large angular scales ($l\lesssim 500$),
\begin{eqnarray}
k^2 P(k)\approx l^2 C_l&\approx& (8\pi G \mu )^2\sqrt{\pi}\frac{\mathcal{L}\hat{\xi}}{2 \mathcal{A}}\frac{\overline{v}^2}{\overline{t}} \ln\left( \frac{t_0}{t}\right)\\
&\approx& (8\pi G \mu )^2\sqrt{\pi}\frac{\mathcal{L}\hat{\xi}}{2 \mathcal{A}}\frac{\overline{v}^2}{\overline{t}} \ln\left( \frac{t_0}{t_{lss}}\frac{l}{500}\right)
\end{eqnarray}
This is only strictly valid, as mentioned above, for $l\gtrsim 60$ where the flat sky approximation remains good. However, it can be extrapolated to lower $l$ to provide useful estimates as can our subsequent
results for the bispectrum and trispectrum in these regimes.   From  this expression and the small angle result, we see that the peak of the power spectrum lies near $l\approx 500$ with a logarithmic deviation from scale-invariance on large angular scales and a steeper $l^{-1}$ fall-off on smaller scales.  The large angle result is qualitatively in agreement with early numerical work on this problem in ref.~\cite{Pogosian} (see, more recent results in \cite{Battye}), though we note that quantitatively the rise towards a peak at $l\approx 500$ is much
steeper for two reasons: (i) The matter-radiation transition breaks scale-invariance because the
density of strings is substantially higher in the radiation era and is therefore falling through recombination
and afterwards as the string slowly responds to the smooth increase in the expansion rate.  (ii) We do not take into account  the pre-recombination perturbations seeded in the
cosmological fluid by cosmic strings which create additional intrinsic CMB anisotropies.  Since the
power spectrum is not our focus here in this analytic work, we leave these quantitative issues for
further study elsewhere \cite{FergLandetal}.

\section{Bispectrum}
\subsection{Bispectrum on small angular scales}
The calculation above for the power spectrum (\ref{uetcpower}) can be easily extended to the bispectrum by defining
the analogous unequal time correlators
\begin{eqnarray*}
\left<\delta_{\mathbf{k}_1}(t_1)\delta_{\mathbf{k}_2}(t_2)\delta_{\mathbf{k}_3}(t_3)\right>=(2\pi)^2\delta(\mathbf{k_1+k_2+k_3})B^T(\mathbf{k}_1,\mathbf{k}_2,\mathbf{k}_3,t_1,t_2,t_3)\,,
\end{eqnarray*}
which are again found by integrating over the string sources 
\begin{eqnarray*}
&&B^T(\mathbf{k}_1,\mathbf{k}_2,\mathbf{k}_3,t_1,t_2,t_3)=\\
&&i(8\pi G \mu )^3\frac{1}{\mathcal{A}}\frac{k_1^A k_2^B k_3^C}{k_1^2 k_2^2 k_3^2}\int d\sigma_1d\sigma_2d\sigma_3\left<	\dot{X}^A(\sigma_1,t_1)\dot{X}^B(\sigma_2,t_2)\dot{X}^C(\sigma_3,t_3)	e^{i\mathbf{k_a.X_a}     }\right>
\end{eqnarray*}
where $a=1,2,3$.   As for the power spectrum, 
the bispectrum is found by integration of the unequal time bispectrum between last scattering and today  $t_i\in [t_{lss},t_0]  $.  A similar calculation to this was carried out in \cite{Hindmarsh2009}, except that 
the contribution was considered to be dominated by the strings near the last scattering surface i.e. 
$t\approx t_{lss}$.   Just for clarity at the outset we present some of our notation.   We are concerned
with triangles formed by the three wavevectors ${\bf k}_1,\,{\bf k}_2,\,{\bf k}_3$ such that ${\bf k}_1+{\bf k}_2+{\bf k}_3=0$ so we define
\begin{eqnarray*}
k_1 = |{\bf k}_1|\,,\quad k_2 = |{\bf k}_2|\,,\quad k_3 = |{\bf k}_3|\,,\qquad \kappa_{12} = {\bf k}_1\cdot {\bf k}_2\,,\quad  \kappa_{23} = {\bf k}_2\cdot {\bf k}_3\,,\quad  \kappa_{13} = {\bf k}_1\cdot {\bf k}_3\,.
\end{eqnarray*}
A particular concern will be the area of the triangle (especially for flattened cases where it vanishes) with
\begin{eqnarray*}
&&{\rm Area}_\Delta = \frac{1}{4} \sqrt{2(k_1^2k_2^2 + k_2^2k_3^2 + k_3^2k_1^2) - k_1^4 -k_2^4-k_3^4    } =\frac{1}{2}  \sqrt {k_1^2 k_2^2 - \kappa_{12}}  ...  \cr
&& K_1 = 2 {\rm Area}_\Delta/k_1\,, \quad K_2 = 2 {\rm Area}_\Delta/k_2\,, \quad K_3 = 2 {\rm Area}_\Delta/k_3\,.
\end{eqnarray*}
We note also that we will temporarily drop the subscript referring to time, which we will treat 
as dimensionless with the range of $t$ in $[1,t_0/t_{lss}]$.   We will continue to use $k$ in place
of $l$, but their relation is given by (\ref{wavemultipole})s in the flat sky approximation.  

Writing $C^{ABC}=\dot{X}^A(\sigma_1)\dot{X}^B(\sigma_2)\dot{X}^C(\sigma_3)	$ and $D=k_a.X_a$, and  again using the assumption of a Gaussian process we find that
\begin{eqnarray*}
\left<	C^{ABC}e^{iD}	\right>=i\left(\left<	C^{ABC}D	\right>    -\left<\dot{X}^A(\sigma_1)D\right> \left<\dot{X}^B(\sigma_2)D\right>\left<\dot{X}^C(\sigma_3)D\right>   \right)e^{-\frac{D^2}{2}}
\end{eqnarray*}
Now writing $\dot{X}_1=\dot{X}(\sigma_1)$, etc we have
\begin{eqnarray*}
\left<	C^{ABC}D	\right>&=&\left< 	\dot{X}_1^A	\dot{X}_2^B\right>\left< 	\dot{X}_3^C	\mathbf{k_a.X_a}\right>+\left< 	\dot{X}_1^A	\dot{X}_3^C\right>\left< 	\dot{X}_2^B	\mathbf{k_a.X_a}\right>\\
&&+\left< 	\dot{X}_2^B	\dot{X}_3^C\right>\left< 	\dot{X}_1^A	\mathbf{k_a.X_a}\right>\\
&=&\frac{\delta^{AB}}{2}V(\sigma_1-\sigma_2)\left[\left<	\dot{X}_3^C k_1^D(X_1^D-X_3^D)+\dot{X}_3^C k_2^D(X_2^D-X_3^D)	\right>	\right]+\dots\\
&=&\frac{\delta^{AB}}{4}V(\sigma_1-\sigma_2)\left[k_1^C\Pi(\sigma_1-\sigma_3)+k_2^C\Pi(\sigma_2-\sigma_3)	\right]\\
&&+\frac{\delta^{AC}}{4}V(\sigma_1-\sigma_3)\left[k_1^B\Pi(\sigma_1-\sigma_2)+k_3^B\Pi(\sigma_3-\sigma_2)	\right]\\
&&+\frac{\delta^{BC}}{4}V(\sigma_2-\sigma_3)\left[k_2^A\Pi(\sigma_2-\sigma_1)+k_3^A\Pi(\sigma_3-\sigma_1)	\right]
\end{eqnarray*}
and
\begin{eqnarray*}
\left<	D^2	\right>&=&\left<\mathbf{k_a.X_a}	\mathbf{k_b.X_b}\right>\\&=&\left<(k_1^A(X_1^A-X_3^A)+k_2^A(X_2^A-X_3^A))(k_2^B(X_2^B-X_1^B)+k_3^B(X_3^B-X_1^B))\right>\\
&=&-\frac{1}{2}\left[	\kappa_{13}\Gamma(\sigma_1-\sigma_3)+\kappa_{23}\Gamma(\sigma_2-\sigma_3)	+\kappa_{12}\Gamma(\sigma_1-\sigma_2)		\right]
\end{eqnarray*}
where $\kappa_{12}=\mathbf{k_1 . k_2}$, etc. We also find that
\begin{eqnarray*}
&&\left<\dot{X}_i^A D\right>=\sum_{j\neq i}\frac{k_j^A}{2} \Pi_{ji}\quad \implies\\
&& k_1^A k_2^B k_3^C \left<\dot{X}^A_1D\right>  \left<\dot{X}^B_2 D\right> \left<\dot{X}^C_3 D\right>=\\ &&\qquad\qquad\qquad \frac{1}{8}(\kappa_{12}\Pi_{12}+\kappa_{13}\Pi_{13})(\kappa_{12}\Pi_{12}+\kappa_{23}\Pi_{23})(\kappa_{13}\Pi_{13}+\kappa_{23}\Pi_{23})
\end{eqnarray*}
(where we use the notation $\Pi_{ij}=\Pi(\sigma_i-\sigma_j)$). The second term is subdominant (as in the case of the power spectrum) and so we neglect it in the following.
\par
In order to calculate this we note that only two of $\sigma_{12},\sigma_{13},\sigma_{23}$ are independent. The choice of independent variables depend on the quantities under consideration, e.g. for the term $V_{13}\Pi_{12}$ we use $\sigma_{12}, \sigma_{13}$ and any term involving $\sigma_{23}$ is then written in terms of these variables. Therefore the prescription is to make a transformation from $\sigma_1,\sigma_2,\sigma_3$ to these variables. Clearly one of the integrations is then independent of the two independent variables chosen and integrates to give $\mathcal{L}$, as in the case of the power spectrum. In the following we consider the term $V_{13}\Pi_{12}$, i.e.
\begin{eqnarray*}
\int d\sigma_1 d\sigma_2 d\sigma _3 V_{13}\Pi_{12}\exp(-D^2/2)=\frac{\mathcal{L}}{4}\int d\sigma_{12}d\sigma_{13} V_{13}\Pi_{12}\exp(-D^2/2).
\end{eqnarray*}
We can assume the small angle approximations for $\Gamma, \Pi$ but as we will outline we need to use the better approximation in the case of $V$, as detailed in Section $3$. In order to simplify the calculation we note that
\begin{eqnarray*}
\frac{-D^2}{2}&\approx&\frac{\overline{t}^2}{4}\left(	\kappa_{12}\sigma_{12}^2+ \kappa_{13}\sigma_{13}^2 +\kappa_{23}	(\sigma_{12}-\sigma_{13})^2	\right)\\
&\approx&-\frac{\overline{t}^2}{4}\left((k_2\sigma_{12}-\frac{\kappa_{23}}{k_2}\sigma_{13})^2  +\frac{k_2^2 k_3^2 -\kappa_{23}^2}{k_2^2}\sigma_{13}^2	\right).
\end{eqnarray*}
This indicates that we can make the integration separable using
\begin{eqnarray*}
&&\int d\sigma_{12}d\sigma_{13} V_{13}\Pi_{12}\exp(-D^2/2)\\
&&\approx\int d\sigma_{12} \Pi_{12}\exp\left(-\frac{\overline{t}^2 k_2^2 \sigma_{12}^2}{4}\right)\int d\sigma_{13}V_{13}\exp\left(-\overline{t}^2 \frac{ \mbox{Area}_{\triangle}^2}{k_2^2} \sigma_{13}^2\right)\\
&&\approx\int d\sigma_{12} \Pi_{12}\exp\left(-\frac{\overline{t}^2 k_2^2 \sigma_{12}^2}{4}\right)\int d\sigma_{13}V_{13}\exp\left(-\frac{\overline{t}^2 K_2^2 \sigma_{13}^2}{4}\right)
\end{eqnarray*}
where $\mbox{Area}_{\triangle}=\sqrt{k_2^2 k_3^2 -\kappa_{23}^2}/2$ and (for simplicity of notation) we denote $K_i= 2 \mbox{Area}_{\triangle}/k_i $. Since $K_i$ (divided by $t_0$) may be less than $500$, i.e. correspond to an angular scale below that of the correlation length then we should use a better approximation to $V_{12}$. From earlier considerations we find that
\begin{eqnarray*}
&&\int d\sigma \Pi(\sigma)\exp\left(-\frac{\overline{t}^2 k^2 \sigma^2}{4}\right)=\frac{c_0}{\hat{\xi}}\int_{0}^{\hat{\xi}/2} d\sigma \sigma^2 \exp\left(-\frac{\overline{t}^2 k^2 \sigma^2}{4}\right)\\
&&\approx \frac{c_0}{\hat{\xi}}\frac{2\sqrt{\pi}\mbox{erf}(k\hat{\xi}\overline{t}/4)-k\hat{\xi}\overline{t}\exp(-(k\hat{\xi}\overline{t}/4)^2)}{k^3 \overline{t}^3}\equiv \frac{c_0}{\hat{\xi}}\frac{1}{k^3 \overline{t}^3}f_1(k\hat{\xi})
\end{eqnarray*}
and also
\begin{eqnarray*}
&&\int d\sigma V(\sigma)\exp\left(-\frac{\overline{t}^2 K^2 \sigma^2}{4}\right)\approx 2\overline{v}^2 \int_0^{\infty} d\sigma \left(1-\frac{\sigma}{\hat{\xi}}\right)\exp\left(-\sigma/\hat{\xi}\right)\exp\left(-\frac{\overline{t}^2 K^2 \sigma^2}{4}\right)\\
&&\approx 2\overline{v}^2 \frac{\sqrt{\pi}(2+\hat{\xi}^2K^2\overline{t}^2) \mbox{erfc}(1/(\hat{\xi}K\overline{t})) \exp(1/(\hat{\xi}K\overline{t}))^2-2\hat{\xi}K\overline{t}}{K^3 \hat{\xi}^2\overline{t}^3}\equiv\frac{2\overline{v}^2}{K^3 \hat{\xi}^2 \overline{t}^3}f_2(K\hat{\xi})
\end{eqnarray*}
Using this notation (i.e. the above definitions of $f_1$ and $f_2$) the total bispectrum on small angular scales (for which all $k_i\gtrsim 500$), after some algebra, reads
\begin{eqnarray}\label{bispectrumfinal}
B({k_1,k_2,k_3})=&&\int_1^{t_0/t_{lss}} dt(8\pi G\mu)^3 \frac{\mathcal{L}\hat{\xi}}{\mathcal{A}}\frac{c_0 \overline{v}^2}{64 \overline{t}^6 \hat{\xi}^4}\frac{1}{k_1^2 k_2^2 k_3^2 \mbox{Area}_{\triangle}^3}~\times\\
&&\left( \kappa_{12}k_3^2 f_1(k_3\hat{\xi})f_2(K_3 \hat{\xi}) +\kappa_{13}k_2^2f_1(k_2\hat{\xi})f_2(K_2 \hat{\xi}) +\kappa_{23}k_1^2 f_1(k_1\hat{\xi})f_2(K_1 \hat{\xi}) \right)
\end{eqnarray}
In order to carry out the time integration we identify $k_i, K_i$ as angular multipoles and $\hat{\xi}=\tilde{\xi}t$ (where we use the renormalised time as usual).   We have calculated the bispectrum numerically 
over the full range of multipoles for which the flat sky approximation is valid and the results are shown 
in Figures 3 and 4.    The end result is a fairly featureless flat bispectrum with very localised and 
modest upturns for flattened triangles and suppression of squeezed triangles because of causality. 
This is quite unlike the inflationary bispectra found in ref.~\cite{Fergusson} and should be easily 
distinguishable given a sufficiently significant signal.   In the analytic spirit of this paper, however, 
we press on to give some simple analytic approximations to the full bispectrum (\ref{bipsectrumfinal})
in all the different regimes:

\paragraph{Approximation (1):}
In the small angle limit at high multipoles with  $k_i\hat{\xi}, K_i\hat{\xi}\gtrsim 10$ at $t_{lss}$ we have $f_1(k_i\hat{\xi})\approx 2\sqrt{\pi}$ and $f_2(K_i \hat{\xi})\approx \sqrt{\pi}\hat{\xi}^2 K_i^2 \overline{t}^2$. This then implies that
\begin{eqnarray*}
B({k_1,k_2,k_3})&\approx&-\int_1^{t_0/t_{lss}} dt(8\pi G\mu)^3 \frac{\mathcal{L}\hat{\xi}}{\mathcal{A}}\frac{c_0 \overline{v}^2 \pi}{64 \overline{t}^4 \hat{\xi}^2}\frac{k_1^2+k_2^2+k_3^2}{k_1^2 k_2^2 k_3^2 \mbox{Area}_{\triangle}}\\
&\approx&-(8\pi G\mu)^3 \frac{\mathcal{L}\hat{\xi}}{\mathcal{A}}\frac{c_0 \overline{v}^2\pi}{64 \overline{t}^4 \tilde{\xi}^2}\frac{k_1^2+k_2^2+k_3^2}{k_1^2 k_2^2 k_3^2 \mbox{Area}_{\triangle}}\\
\implies (k_1^2 k_2^2k_3^2)^{2/3}B(k_1,k_2,k_3)&\approx&-(8\pi G\mu)^3 \frac{\mathcal{L}\hat{\xi}}{\mathcal{A}}\frac{c_0 \overline{v}^2\pi}{64 \overline{t}^4 \tilde{\xi}^2}\frac{k_1^2+k_2^2+k_3^2}{(k_1^2 k_2^2 k_3^2)^{1/3} \mbox{Area}_{\triangle}}\propto \frac{1}{\tilde{\xi}^2 l^2}
\end{eqnarray*}
The difference between this result and that of ref.~(\cite{Hindmarsh2009}) is due to the approximation made to make the integration over the $\sigma$ coordinates separable. This has little effect on the quantitative result and we observe the same asymptotic limit.

\paragraph{Approximation (2):}
For $K\hat{\xi}\gtrsim 5$ (at the initial time $t_{start}$) and any value of $k$ we find that 
\begin{eqnarray*}
&& (k_1^2 k_2^2k_3^2)^{2/3}B(k_1,k_2,k_3)\approx (8\pi G\mu)^3 \frac{\mathcal{L}\hat{\xi}}{\mathcal{A}}\frac{c_0 \overline{v}^2 \pi}{64 \overline{t}^4 \tilde{\xi}^2}\\
&&\frac{5(\kappa_{23} \mbox{erf}(k_1\hat{\xi}\overline{t}/4)+\kappa_{13}  \mbox{erf}(k_2\hat{\xi}\overline{t}/4)+\kappa_{12}  \mbox{erf}(k_3\hat{\xi}\overline{t}/4))}{3(k_1^2 k_2^2 k_3^2)^{1/3} \mbox{Area}_{\triangle}}
 \end{eqnarray*}
where $k_m=\min(k_i,500)$ tells us the earliest time, $t_{start}$, to which we can integrate back and for which $k_m\hat{\xi}=(k_m/500)t/t_{lss} =1$ at $t=t_{start}$.

\paragraph{Approximation (3):}
For the intermediate range $K\hat{\xi}\in (3,5)$ we find 
\begin{eqnarray*}
&& (k_1^2 k_2^2k_3^2)^{2/3}B(k_1,k_2,k_3)\approx(8\pi G\mu)^3 \frac{\mathcal{L}\hat{\xi}}{\mathcal{A}}\frac{c_0 \overline{v}^2 \pi}{64 \overline{t}^4 \tilde{\xi}^2}\\
&&\frac{\sqrt{5}(\kappa_{23}  \mbox{erf}(k_1\tilde{\xi}\overline{t}/4)\sqrt{K_1\hat{\xi}}+\kappa_{13}  \mbox{erf}(k_2\hat{\xi}\overline{t}/4)\sqrt{K_2\tilde{\xi}}+\kappa_{12} \mbox{erf}(k_3\hat{\xi}\overline{t}/4)\sqrt{K_3\tilde{\xi}})}{3(k_1^2 k_2^2 k_3^2)^{1/3} \mbox{Area}_{\triangle}}
 \end{eqnarray*}
This approximation is adjusted slightly for $K\hat{\xi}\in (1,3)$ where we replace $\mbox{erf}(k_3\hat{\xi}\overline{t}/4)$ by $0.8\mbox{erf}(k_3\hat{\xi}\overline{t}/2.5)$. However since we are interested in only a quantitative approximation to the level of signal from the bispectrum we note that we can use the expression for $K\hat{\xi}\in (3,5)$ in the full range $(1,5)$.

\paragraph{Approximation (4):}
For flattened triangles $K\hat{\xi}\lesssim 1$ the behaviour of the bispectrum gives (for small angles $k\hat{\xi}\gtrsim 4$) we need an improved estimate finding
\begin{eqnarray*}
&& (k_1^2 k_2^2k_3^2)^{2/3}B(k_1,k_2,k_3)\approx (8\pi G\mu)^3 \frac{\mathcal{L}\hat{\xi}}{\mathcal{A}}\frac{c_0 \overline{v}^2 \pi}{64 \overline{t}^4 \tilde{\xi}^2}\\
&&\frac{\sqrt{5}(\kappa_{23}  \mbox{erf}(k_1\hat{\xi}\overline{t}/4){K_1\tilde{\xi}}+\kappa_{13}   \mbox{erf}(k_2\hat{\xi}\overline{t}/4){K_2\tilde{\xi}}+\kappa_{12}  \mbox{erf}(k_3\hat{\xi}\overline{t}/4){K_3\tilde{\xi}})}{6(k_1^2 k_2^2 k_3^2)^{1/3} \mbox{Area}_{\triangle}}\\
&&\approx(8\pi G\mu)^3 \frac{\mathcal{L}\hat{\xi}}{\mathcal{A}}\frac{c_0 \overline{v}^2 \pi}{32 \overline{t}^4 \tilde{\xi}}\frac{\kappa_{23} \mbox{erf}(k_1\hat{\xi}\overline{t}/4)+\kappa_{13} \mbox{erf}(k_2\hat{\xi}\overline{t}/4)+\kappa_{12} \mbox{erf}(k_3\hat{\xi}\overline{t}/4)}{3(k_1^2 k_2^2 k_3^2)^{1/3}}\\
 \end{eqnarray*}
 This approximation is further motivated by the asymptotic expansion of $\mbox{erfc}(x)\approx \exp(-x^2)\frac{1}{\sqrt{\pi}x}+\dots$ for $x$ large.
 For $k\hat{\xi}<4$ we note that the formula behaves more like $Approx.~(3)$ for lower and lower values of $K\hat{\xi}$ before it becomes accurate. It is important from these approximations to note that although the bispectrum grows as we approach collapsed configurations (for which $\mbox{Area}_{\triangle}\rightarrow 0$) it is not divergent and attains a finite value on the boundary.

\medskip
To summarise, let us provide a final simple expression which is valid over all multipoles
$l<2000$ (e.g.\ relevant for both the Planck and WMAP experiments):
\begin{eqnarray} \label{Planckbispectrum}
&& (k_1^2 k_2^2k_3^2)^{2/3}B(k_1,k_2,k_3)\approx (8\pi G\mu)^3 \frac{\mathcal{L}\hat{\xi}}{\mathcal{A}}\frac{c_0 \overline{v}^2 \pi}{64 \overline{t}^4 \tilde{\xi}^2}~\times\\
&&\frac{5(\kappa_{23}  \mbox{erf}(k_1\hat{\xi}\overline{t}/4)D(K_1\tilde{\xi})+\kappa_{13}   \mbox{erf}(k_2\hat{\xi}\overline{t}/4)D(K_2\tilde{\xi})+\kappa_{12}  \mbox{erf}(k_3\hat{\xi}\overline{t}/4)D(K_3\tilde{\xi}))}{3(k_1^2 k_2^2 k_3^2)^{1/3} \mbox{Area}_{\triangle}}
\end{eqnarray}
where the function $D$ is defined by 
\begin{eqnarray}\label{Planckfunctionone}
D(K\tilde{\xi})&=&1\qquad\qquad \quad\mbox{if} \,\, K\tilde{\xi}>5 \\\label{Planckfunctiontwo}
&=&\sqrt{K\tilde{\xi}/5}\qquad \mbox{if} \,\, K\tilde{\xi}\in(0,5)\,..
\end{eqnarray}
Although this expression will not accurately describe the very localised upturn near the edges of tetrahedron, this region makes a negligible integrated contribution to $f_{NL}$.   Going further 
out beyond $l>2000$, these upturned edge regions become more significant, so we must replace the 
second line above (\ref{Planckfunctiontwo}) by the following
\begin{eqnarray*}
D(K\tilde{\xi})&=&\sqrt{K\tilde{\xi}/5}\qquad \mbox{if} \,\, K\tilde{\xi}\in(1,5)\,,\\
&=&K\tilde{\xi}/\sqrt{5}\qquad \mbox{if} \,\, K\tilde{\xi}<1\,..
\end{eqnarray*}

\subsection{Bispectrum on large angular scales}
If $\min k_i =k_m\lesssim 500$, then we must restrict the range of times over which we sum to $(t_{start},t_0)$. We find that, as in the case for the power spectrum, the bispectrum decreases as a logarithm as we restrict the range, i.e. the bispectrum $\propto \ln((k_m/500) t_0/t_{lss}) \approx \ln(k_m/10)$. Therefore, we can include this effect by multiplying the formulae in the previous section by $\ln(k_m/10)/\ln(50)$ for $k_m\lesssim 500$. The extension to large angular scales allows us to make approximations to the level of non-Gaussianity in the parameter range of Planck.
In Figures 5 and 6 we illustrate $\tilde k = k_1+k_2+k_3 = const.$ slices through the 
bispectrum to reveal the cross-sectional shape using the approximations above.

\subsection{Estimator of $f_{NL}$}
\begin{figure}[htp]
\centering 
\includegraphics[width=152mm]{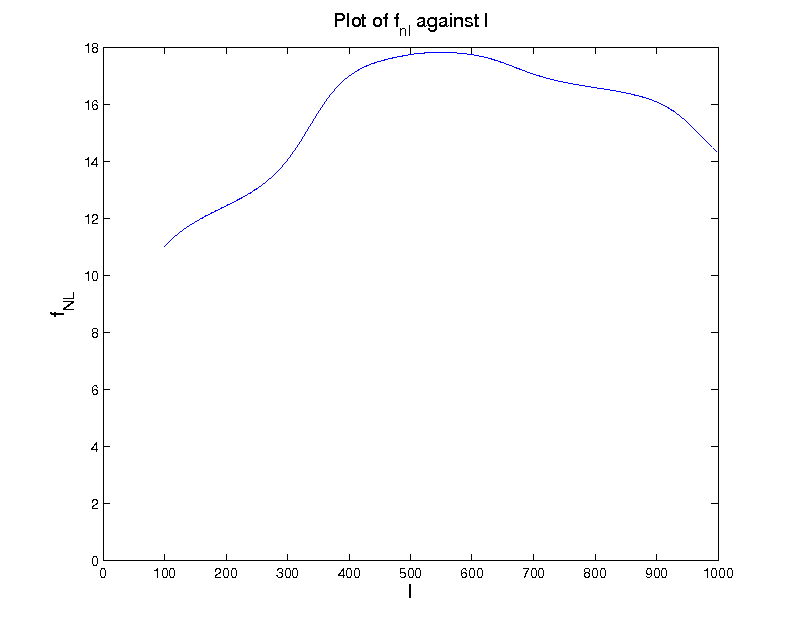}
\caption{Measure of the level of non-Gaussianity using $f_{NL}^{\mbox{local}}$}
\label{fig:fnl}
\end{figure}

As detailed in \cite{babich} the signal to noise ratio of the bispectrum is given by
\begin{eqnarray*}
(S/N)^2=\frac{1}{4\pi}\sum_{l_1, l_2 ,l_3}(2 l_1+1)(2 l_2+1)(2 l_3+1)\left( \begin{array} {ccc}
l_1 & l_2 &l_3 \\
0 & 0  &0\end{array} \right)^2 \frac{b_{l_1,l_2,l_3}^2}{C_{l_1}C_{l_2}C_{l_3}}
\end{eqnarray*}
where $b_{l_1 l_2 l_3}$ corresponds to the reduced bispectrum, which is precisely the bispectrum as calculated in the flat sky limit and where we assume that the noise is cosmic variance dominated.
From the images shown in the last section we see that in the range of interest for Planck we can find a reasonably accurate estimate for $f_{NL}$ by summing over the equal $l$ values. This gives
\begin{eqnarray*}
(S/N)^2&\approx&\frac{2}{\pi}\sum_{l}l^4 \left( \begin{array}{ccc}
l & l &l \\
0 & 0  &0\end{array} \right)^2 \frac{b_{l_1,l_2,l_3}^2}{C_{l_1}C_{l_2}C_{l_3}}\\
&\approx&\frac{1}{2\sqrt{3}\pi^5}\sum_{l\in 2\mathcal{Z}} \frac{l (l^4 b_{lll})^2}{(l^2 C_l/(2\pi))^3}
\end{eqnarray*}
For the local model of inflation the signal to noise ratio is proportional to $f_{NL}$. Since \cite{Fergusson} the local bispectrum is proportional to $l^{-4}$ for $l\lesssim 1000$ we normalise the signal to noise ratio of cosmic strings to that of local models of inflation, for which $l^4 b_{lll}\approx 2\times 10^{-17}f_{NL}$. We have computed $l^2 C_l/(2\pi)$ using CMBFAST with WMAP5 values for the parameters.
Therefore, defining,
\begin{eqnarray*}
f_{NL}^2=(S/N)^{2(\mbox{cosmic strings})}/(S/N)^{2(\mbox{inflation})},
\end{eqnarray*}
we plot $f_{NL}$ against $l$ for multipoles summed between 60 and $l$. The fluctuations in the plot arise since cosmic strings do not strictly follow a local model and because of the transfer functions appearing in the $C_l$s. However, we find that for the following values of the cosmic string model parameters, as found in simulations (\cite{BBS},\cite{MartShell}, \cite{AllShel})
\begin{eqnarray*}
\overline{v}^2=0.18;\qquad \overline{t}^2=0.42; \qquad \frac{\mathcal{L}\hat{\xi}}{\mathcal{A}}=2
\end{eqnarray*}
and $G\mu=7\times 10^{-7}$ that $f_{NL}\approx -75 c_0$. Preliminary estimates of $c_0$ from numerical simultaions suggest that it is significantly below unity, so we estimate $f_{NL}\sim -20$ (see Figure ~\ref{fig:fnl}). Given the most recent estimate of $f_{NL}$ for local models $f_{NL}=38\pm 21$ (\cite{Zaldarr} ). It should also be stressed that cosmic strings produce a very different shape to  local non-Gaussianity, so analysis of compatibility with WMAP or Planck needs to be specifically investigated (\cite{FergLandetal}).

\subsection{Numerical results}
\begin{figure}[htp]
\centering 
\includegraphics[width=102mm]{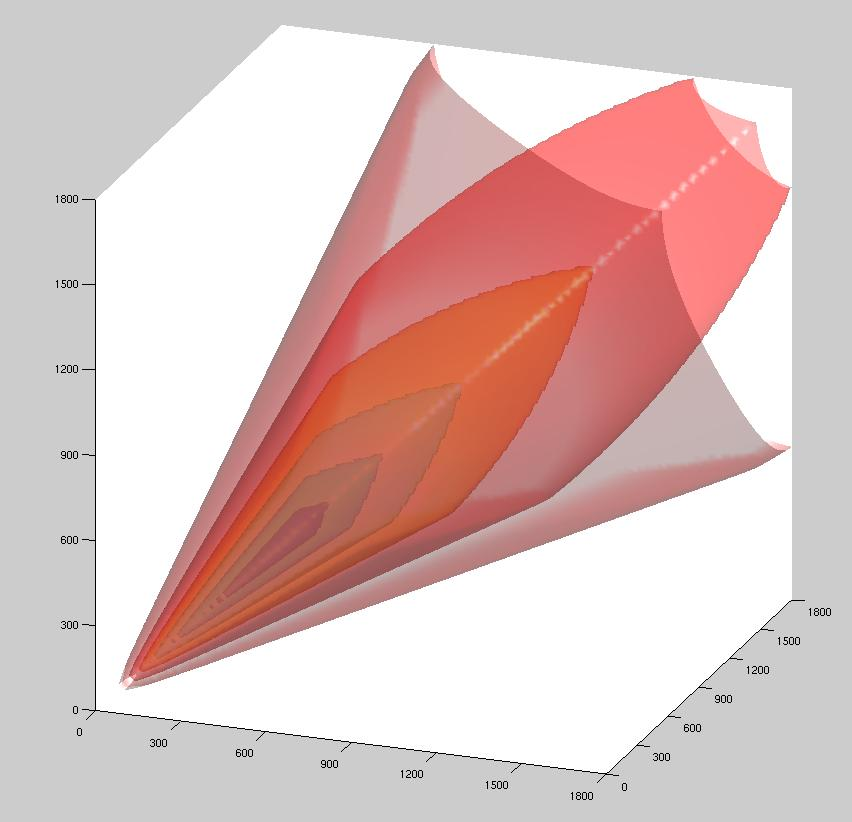}
\caption{Plot of the 3D bispectrum  $(l_1 l_2 l_3)^{4/3}B(l_1,l_2,l_3)$. The signal is seen to peak at the near the correlation length at last scattering for which all $l_i\approx 500$. Due to the resolution of the data points we do not pick up the rise towards the edge.}
\label{fig:bispec1}
\end{figure}

\begin{figure}[htp]
\centering 
\includegraphics[width=102mm]{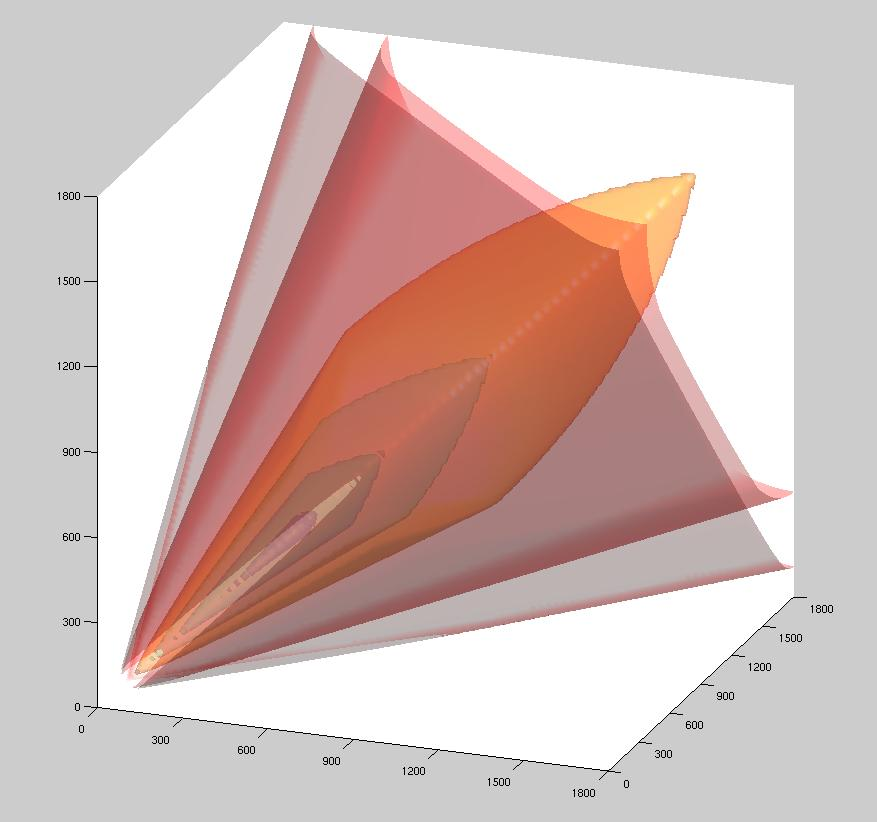}
\caption{Plot of the 3D bispectrum  $(l_1 l_2 l_3)^{4/3}B(l_1,l_2,l_3)$ with difference slicings to that of Figure~\ref{fig:bispec1}.}
\label{fig:bispec2}
\end{figure}

\begin{figure}[htp]
\centering 
\includegraphics[width=102mm]{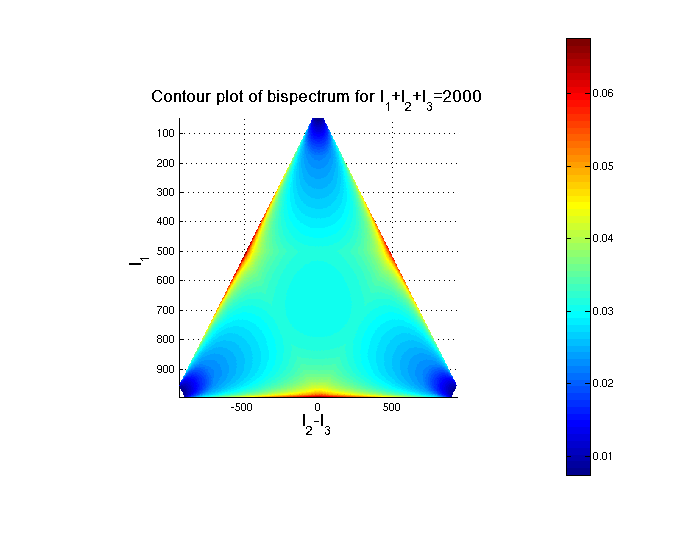}
\caption{Contour plot of the slice through the bispectrum at $\sum_i l_i=2000$.}
\label{fig:slice2000}
\end{figure}

\begin{figure}[htp]
\centering 
\includegraphics[width=102mm]{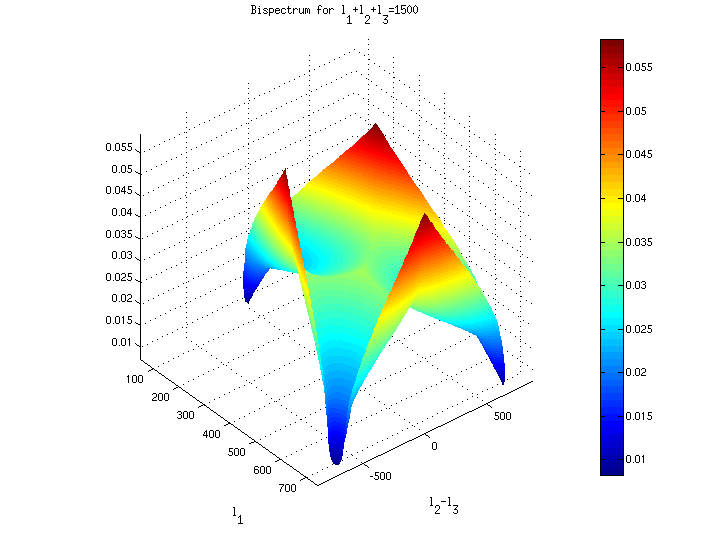}
\caption{Plot of the slice through the bispectrum at $\sum_i l_i=1500$.}
\label{fig:slice1500}
\end{figure}

We have plotted the three dimensional bispectrum in the multipole range of Planck $l_i< 2000$ by numerically integrating the analytic result (see Figures~\ref{fig:bispec1} and~\ref{fig:bispec2} ). In particular we plot the quantity $(l_1 l_2 l_3)^{4/3}B(l_1,l_2,l_3)$. We do not integrate out fully to the edge since the result begins to diverge there and is more difficult to evaluate numerically. We find that the signal peaks at around $(500,500,500)$ and drops towards the corners. Therefore, our assumption that using the equal $l$ values to estimate the level of non-Gaussianity seems valid.
However, we do expect that the signal rises very near the edge (before it levels out) and so we use our approximations as detailed earlier to track this behaviour on slices across the bispectrum. In particular we show the slice for $l_1+l_2+l_3=2000$ (Figure~\ref{fig:slice2000}) and for $l_1+l_2+l_3=1500$ (Figure~\ref{fig:slice1500}). In order to see the behaviour more clearly we have multplied the bispectrum by a negative sign and we have shown the former plot as a contour lot. The slices are plotted in units of $\epsilon^3=(8\pi G\mu)^3$. There is a peak towards the edge but only a factor of a few that at the centre point. Also it begins to diverge very near the edge for these slices. Therefore, we feel justified in neglecting the edges in our estimator for $f_{NL}$. 


\section{Trispectrum}
\subsection{Trispectrum on small angular scales}
As in the section on the bispectrum we regard $k$ as the angular multipole and $\hat{\xi}$ to be the correlation length as measured at the original time (normalised by some factor as discussed previously). We will return the to the time dependence later in the calculation. However, we note here that as in the case for the power spectrum and bispectrum the total trispectrum reduces to the sum of equal time four point correlators between $t_{start}\gtrsim t_{lss}$ and $t_0$.
\par
The trispectrum, in the flat sky approximation, is given (at any given time) by
\begin{eqnarray*}
\left<\delta_{\mathbf{k}_1}\delta_{\mathbf{k}_2}\delta_{\mathbf{k}_3}\delta_{\mathbf{k}_4}\right>_c=(2\pi)^2 \delta^{(2)}(\mathbf{k_1+k_2+k_3+k_4})T_c(\mathbf{k_1,k_2,k_3,k_4})
\end{eqnarray*}
where the subscript $c$ denotes the connected part. 
 This suggest that we normalise the trispectrum by the area factor $\mathcal{A}$. However, as we shall see below we should in fact normalise by $\mathcal{A}^2$ for parallelogram configurations. The unconnected part, denoted hereafter by the subscript $uc$, is given by
\begin{eqnarray*}
\left<\delta_{\mathbf{k}_1}\delta_{\mathbf{k}_2}\delta_{\mathbf{k}_3}\delta_{\mathbf{k}_4}\right>_{uc}&=&(2\pi)^4  \delta^{(2)}(\mathbf{k_1+k_2})\delta^{(2)}(\mathbf{k_3+k_4})P(k_1)P(k_3)+(2\pi)^4  \delta^{(2)}(\mathbf{k_1+k_3})\delta^{(2)}(\mathbf{k_2+k_4})P(k_1)P(k_2)\\
&&+(2\pi)^4  \delta^{(2)}(\mathbf{k_1+k_4})\delta^{(2)}(\mathbf{k_2+k_3})P(k_1)P(k_2)\\
&=&\mathcal{A}^2\left( \delta_{k_1,k_2}\delta_{k_3,k_4}P(k_1)P(k_3)+\delta_{k_1,k_3}\delta_{k_2,k_4}P(k_1)P(k_2)+\delta_{k_1,k_3}\delta_{k_2,k_4}P(k_1)P(k_2)\right)
\end{eqnarray*}
We will calculate the four point correlator and subtract off this unconnected part to find the trispectrum. 
\par
In this paper we will consider only parallelogram configurations. In this regard we are motivated by the earlier section on the bispectrum which showed that as the triangle collapses it grows (though does not become infinite). Clearly, the parallelogram shape should replicate this behaviour.
\par
For planar parallelograms (planar shapes are, of course, appropriate since we are using the flat sky approximation) we have that opposite sides are antiparallel. To impose this we need to use two delta functions and this implies that the trispectrum is normalised by $\mathcal{A}^2$. This is true in general, however, which becomes more apparent also when we note that we can decompose the delta via
\begin{eqnarray*}
\delta^{(2)}(\mathbf{k_1+k_2+k_3+k_4})=\int d^2\mathbf{L}\delta^{(2)}(\mathbf{k_1+k_2+L})\delta^{(2)}(\mathbf{k_1+k_2-L}).
\end{eqnarray*}
Then we infer that we can approximate the total trispectrum by summing over the possible parallelogram configurations (to specify a parallelogram configuration it suffices to have two of the sides and a diagonal)\footnote{We shall see later in the section that the Gaussian contribution is given by square configurations.}. 
\par
Motivated by the above discussion we define
\begin{eqnarray*}
T^T(\mathbf{k}_1,\mathbf{k}_2,\mathbf{k}_3,\mathbf{k}_4)=\frac{1}{\mathcal{A}^2}\left<\delta_{\mathbf{k}_1}\delta_{\mathbf{k}_2}\delta_{\mathbf{k}_3}\delta_{\mathbf{k}_4}\right>
\end{eqnarray*}
where the multipoles trace out a parallelogram configuration.
\par
In a similar fashion to the calculation of the power spectrum and bispectrum, we find that
\begin{eqnarray*}
T^T(\mathbf{k}_1,\mathbf{k}_2,\mathbf{k}_3,\mathbf{k}_4)=&&\frac{(8\pi G \mu)^4}{\mathcal{A}^2}\frac{k_1^A k_2^B k_3^C k_4^D}{k_1^2 k_2^2 k_3^2 k_4^2}\int d\sigma_1 d\sigma_2 d\sigma_3 d\sigma_4 \left< \dot{X}^A(\sigma_1)\dot{X}^B(\sigma_2)\dot{X}^C(\sigma_3)	\dot{X}^D(\sigma_4)	e^{i\mathbf{k_a.X_a}}   \right>.
\end{eqnarray*}
Writing $C^{ABCD}=\dot{X}^A(\sigma_1)\dot{X}^B(\sigma_2)\dot{X}^C(\sigma_3)\dot{X}^D(\sigma_4)	$ and $E=k_a.X_a$ and using again the assumption of a Gaussian process we get that
\begin{eqnarray*}
\left<	C^{ABCD}e^{iE}	\right>=\left< C^{ABCD}- \frac{1}{2} C^{ABCD} E^2+\frac{1}{24}C^{ABCD}E^4+\dots		\right>
\end{eqnarray*}
Now
\begin{eqnarray*}
\left<	C^{ABCD}	\right>	&=&\left<	\dot{X}^A_1	\dot{X}^B_2	\dot{X}^C_3	\dot{X}^D_4	\right>
=\frac{\delta^{AB}}{2}V_{12}\frac{\delta^{CD}}{2}V_{34}+\frac{\delta^{AC}}{2}V_{13}\frac{\delta^{BD}}{2}V_{24}+\frac{\delta^{AD}}{2}V_{14}\frac{\delta^{BC}}{2}V_{23}\\
\implies k_1^A k_2^B k_3^C k_4^D \left<	C^{ABCD}	\right>&=&\frac{1}{4}\left(\kappa_{12}\kappa_{34}V_{12}V_{34}+\kappa_{13}\kappa_{24}V_{13}V_{24}+\kappa_{14}\kappa_{23}V_{14}V_{23}	\right)
\end{eqnarray*}
Next we consider (writing $C^{ABCD}$ as $ABCD$)
\begin{eqnarray*}
\left<	ABCD E^2\right>=&&\left<	C^{ABCD}	\right>\left<	E^2	\right> +
2\left<AB\right>\left<CE\right>\left<DE\right>+\circlearrowleft
\end{eqnarray*}
where $\circlearrowleft$ denotes the contribution from terms found by permutating the symbols $A,B,C,D$ in the the second term on the right hand side of the equality\footnote{We note that the factor of two appears since $\left<AB\right>=\left<BA\right>$.}.
Now we note that (using $\sum_i \mathbf{k}_i=0$)
\begin{eqnarray*}
\left<AB\right>\left<CE\right>\left<DE\right>&=&\frac{\delta^{AB} V_{12}}{2}\left<\dot{X}^C_3 \sum_{j\neq 3} k_j^E X_{j3}^E\right>\left<\dot{X}^D_4 \sum_{j\neq 4} k_j^E X_{j4}^E\right>\\
&=&\frac{\delta^{AB} V_{12}}{8}\sum_{j\neq 3}k_j^C \Pi_{3j} \sum_{j\neq 4}k_j^D \Pi_{4j}
\end{eqnarray*}
where we use the fact that $\Pi_{ij}=\left< X^A_{ij} \dot{X}_j^A\right>=-<\dot{X}_j^AX^A_{ji}>$.
Suppressing the summation notation we find that
\begin{eqnarray*}
&&k_1^A k_2^B k_3^C k_4^D\left<	ABCD E^2\right>=\left<	C^{ABCD}	\right>\left<	E^2	\right> +\\
&& \frac{1}{4}\left( V_{12}\kappa_{12}	\kappa_{3j}\Pi_{3j} \kappa_{4j}\Pi_{4j}	+V_{13}\kappa_{13}	\kappa_{2j}\Pi_{2j} \kappa_{4j}\Pi_{4j}	+V_{14}\kappa_{14}	\kappa_{2j}\Pi_{2j} \kappa_{3j}\Pi_{3j}	\right)+\\
&&\frac{1}{4}\left( V_{23}\kappa_{23}	\kappa_{1j}\Pi_{1j} \kappa_{4j}\Pi_{4j}	+V_{24}\kappa_{24}	\kappa_{1j}\Pi_{1j} \kappa_{3j}\Pi_{3j}	+V_{34}\kappa_{34}	\kappa_{1j}\Pi_{1j} \kappa_{3j}\Pi_{3j}	\right).
\end{eqnarray*}

Now we proceed to investigate
\begin{eqnarray*}
\left<	ABCD E^4\right>=&&\left<ABCDE^2\right>\left<E^2\right>+24\left<AE\right>\left<BE\right>\left<CE\right>\left<DE\right>.
\end{eqnarray*}
We have seen above that $\left<AE\right>=-\dfrac{1}{2}\sum_{j\neq 1}k_j^A \Pi_{1j}$ so (again suppressing the summation notation) we have
\begin{eqnarray*}
k_1^A k_2^B k_3^C k_4^D\left<AE\right>\left<BE\right>\left<CE\right>\left<DE\right>=\frac{24}{16} \kappa_{1j}\Pi_{1j} \kappa_{2j}\Pi_{2j} \kappa_{3j}\Pi_{3j} \kappa_{4j}\Pi_{4j} 
\end{eqnarray*}
Finally, we collect the terms to find that
\begin{eqnarray*}
k_1^A k_2^B k_3^C k_4^D\left<	C^{ABCD}e^{iE}	\right>&=&[ \frac{1}{4}\left(\kappa_{12}\kappa_{34}V_{12}V_{34}+\kappa_{13}\kappa_{24}V_{13}V_{24}\kappa_{14}\kappa_{23}V_{14}V_{23}	\right)\\
&&- \frac{1}{8}\left( V_{12}\kappa_{12}	\kappa_{3j}\Pi_{3j} \kappa_{4j}\Pi_{4j}	+\circlearrowleft	\right)+\frac{1}{16} 
\kappa_{1j}\Pi_{1j} \kappa_{2j}\Pi_{2j} \kappa_{3j}\Pi_{3j} \kappa_{4j}\Pi_{4j} ]e^{-\left<E^2\right>/2}
\end{eqnarray*}
where we have 
\begin{eqnarray*}
\left<E^2\right>&=&\left<(k_1^A X_{14}^A+k_2^A X_{24}^A+k_3^A X_{34}^A  )(k_2^B X_{21}^B+k_3^B X_{31}^B+k_4^B X_{41}^B)\right>\\
&=&-\frac{1}{2}\left(\kappa_{12}\Gamma_{12}+\kappa_{13}\Gamma_{13}+\kappa_{14}\Gamma_{14}+\kappa_{23}\Gamma_{23}+\kappa_{24}\Gamma_{24}+\kappa_{34}\Gamma_{34}\right)
\end{eqnarray*}
and where we use the notation $\circlearrowleft$ to represent the non-equivalent contributions by permuting the symbols $(1,2,3,4)$ of the term in the brackets.
The dominant term on all angular scales is found to be given by the first term ($\sim C^{ABCD}e^{-E^2/2}$) and, therefore, in what follows we neglect the remaining terms.
\par
For a parallelogram configuration we have $\kappa_{13}=-k_1^2,\, \kappa_{24}=-k_2^2,\, \kappa_{23}=\kappa_{14}=-\kappa_{34}=-\kappa_{12}$. This implies that, in the small angle approximation,
\begin{eqnarray*}
\left<E^2\right>&=&\frac{-\overline{t}^2}{2}\left(\kappa_{12}\sigma_{12}^2-k_{1}^2\Gamma_{13}-\kappa_{12}\sigma_{14}^2-\kappa_{12}\sigma_{23}^2-k_2^2\sigma_{24}^2+\kappa_{12}\sigma_{34}^2\right)
\end{eqnarray*}
 Again the choice of independent variables is dependent on the integrand under consideration. To elucidate this we consider the term
 \begin{eqnarray*}
\int d\sigma_1 d\sigma_2 d\sigma _3 d\sigma_4 V_{13}V_{24}\exp(-E^2/2)=\frac{\mathcal{L}}{8}\int d\sigma_{12}d\sigma_{13}d\sigma_{24} V_{13}\Pi_{12}\exp(-E^2/2)
\end{eqnarray*}
 In a similar manner to the case of the bispectrum we can simplify the calculation by making a separable approximation. In particular we find 
 \begin{eqnarray*}
\exp(-\left<E^2\right>/2)&=&\exp\left(\frac{\overline{t}^2}{4}\left(\kappa_{12}\sigma_{12}^2-k_{1}^2\Gamma_{13}-\kappa_{12}\sigma_{14}^2-\kappa_{12}\sigma_{23}^2-k_2^2\sigma_{24}^2+\kappa_{12}\sigma_{34}^2\right)\right)\\
&=&\exp\left(-\frac{\overline{t}^2}{4}\left((k_1\sigma_{13}-\frac{\kappa_{12}}{k_1}\sigma_{23})^2+\frac{k_1^2 k_2^2 -\kappa_{12}^2}{k_1^2}\sigma_{24}^2)\right)\right)
 \end{eqnarray*}

\begin{eqnarray*}
\int d\sigma_1 d\sigma_2 d\sigma _3 d\sigma_4 V_{13}V_{24}\exp(-E^2/2)=\frac{\mathcal{L}^2}{4}\int d\sigma_{13}  V_{13} \exp\left(-\frac{k_1^2\overline{t}^2\sigma_{13}^2}{4}\right)\int d\sigma_{24} V_{24} \exp\left(-\frac{K_1^2\overline{t}^2\sigma_{24}^2}{4}\right)
\end{eqnarray*}
where $K_i= 2 \mbox{Area}_{\triangle}/k_i $ as defined in the section on the bispectrum (note that the area of the triangle is half that of the parallelogram). Since $K_2$ (divided by $t_0$) may be less than $500$, i.e. correspond to an angular scale below that of the correlation length then we should use a better approximation to $V_{24}$. Following these approximations for all terms and using the results of the previous sections we find that
\begin{eqnarray*}
T^T(\mathbf{k}_1,\mathbf{k}_2,\mathbf{k}_3,\mathbf{k}_4)\approx \frac{\mathcal{L}^2\hat{\xi}^2}{\mathcal{A}^2}    \frac{(8\pi G\mu)^4}{\hat{\xi}^2 k_1^2 k_2^2 k_3^2 k_4^2}\frac{1}{16} \left(\int  d\sigma_{12}d\sigma_{34}2 \kappa_{12}^2 V_{12}V_{34}\exp(-E^2/2)  +\int d\sigma_{13}d\sigma_{24}k_1^2 k_2^2V_{13} V_{24}\exp(-E^2/2)\right)
\end{eqnarray*}
We note also that in the case of a square this gives $P(k_1)P(k_2)$ at each time. This is the Gaussian contribution. Therefore the approximation to the total trispectrum is given by the sum over non-square parallelogram configurations. In order to compute the integral we use the more accurate approximation for $V$. Using the previous expressions for the separate integrals we find
\begin{eqnarray*}
T^T(\mathbf{k}_1,\mathbf{k}_2,\mathbf{k}_3,\mathbf{k}_4)\approx \frac{\mathcal{L}^2\hat{\xi}^2}{\mathcal{A}^2}    \frac{(8\pi G\mu)^4}{\hat{\xi}^2 k_1^2 k_2^2 k_3^2 k_4^2}\frac{\overline{v}^4}{4} \frac{1}{\overline{t}^6\hat{\xi}^4 \mbox{Area}^3 }\left(2\kappa_{12}^2 f_2(K_2\hat{\xi}^2)f_2(k_2\hat{\xi})+k_1^2 k_2^2	f_2(K_1\hat{\xi})f_2(k_1\hat{\xi})\right)
\end{eqnarray*}
where $\mbox{Area}=2\mbox{Area}_{\triangle}$.
\par
This result must be integrated as usual to give the total trispectrum due to cosmic strings between last scattering (or if $k_m<500$ between $t_{start}=t_{lss}500/k_m$) and today. Returning the time dependence using the renormalised time $t/t_{start}$ and understanding the multipoles to refer to angular multipoles we have that the total trispectrum contribution by the parallelogram considered here is
 \begin{eqnarray*}
T^{total}(\mathbf{k}_1,\mathbf{k}_2,\mathbf{k}_3,\mathbf{k}_4)\approx&& \frac{\mathcal{L}^2\hat{\xi}^2}{\mathcal{A}^2}    \frac{(8\pi G\mu)^4}{\tilde{\xi}^2 k_1^2 k_2^2 k_3^2 k_4^2}\frac{\overline{v}^4}{4} \frac{1}{\overline{t}^6\tilde{\xi}^4  \mbox{Area}^3 }\\
&&\int_{1}^{t_0/t_{lss}} dt\frac{1}{t^6}\left(2\kappa_{12}^2 f_2(K_2\tilde{\xi} t)f_2(k_2\tilde{\xi} t)+k_1^2 k_2^2	f_2(K_1\tilde{\xi}t )f_2(k_1\tilde{\xi} t)\right)
\end{eqnarray*}
In order to understand the behaviour for the trispectrum we note the following approximation to the integral given by
\begin{eqnarray*}
T^{total}(\mathbf{k}_1,\mathbf{k}_2,\mathbf{k}_3,\mathbf{k}_4)\approx&& \frac{\mathcal{L}^2\hat{\xi}^2}{\mathcal{A}^2}    \frac{(8\pi G\mu)^4}{\tilde{\xi}^2 k_1^2 k_2^2 k_3^2 k_4^2}\frac{\overline{v}^4}{4} \frac{\pi}{\overline{t}^2 \mbox{Area} }\\
&&(2\kappa_{12}^2   \mbox{erfc}\left(\frac{1}{\tilde{\xi}k_2 \overline{t}}\right)  \exp\left(\frac{1}{(\tilde{\xi}k_2 \overline{t})^2}\right)\mbox{erfc}\left(\frac{1}{\tilde{\xi}K_2 \overline{t}}\right)  \exp\left(\frac{1}{(\tilde{\xi}K_2 \overline{t})^2}\right)\\
&&+k_1^2 k_2^2	 \mbox{erfc}\left(\frac{1}{\tilde{\xi}k_1 \overline{t}}\right)  \exp\left(\frac{1}{(\tilde{\xi}k_1 \overline{t})^2}\right)\mbox{erfc}\left(\frac{1}{\tilde{\xi}K_1 \overline{t}}\right)  \exp\left(\frac{1}{(\tilde{\xi}K_1 \overline{t})^2}\right)).
\end{eqnarray*}
Using the asymptotic expansion of $\mbox{erfc}$ we find that as $K\rightarrow 0$ the formula is independent of $K$, i.e. of $\mbox{Area}$. Therefore, as for the bispectrum, the trispectrum grows towards the edge but is not divergent (i.e. it reaches a cutoff).

\subsection{Trispectrum on large angular scales}
If $\min k_i =k_m\lesssim 500$, then we must restrict the range of times over which we sum to $(t_{start},t_0)$. Again we find that the trispectrum decreases as a logarithm as we restrict the range, i.e. the trispectrum $\propto \ln((k_m/500) t_0/t_{lss}) \approx \ln(k_m/10)$. Therefore, we can include this effect by multiplying the formulae in the previous section by $\ln(k_m/10)/\ln(50)$ for $k_m\lesssim 500$.
This is the same effect as was evident in the case of the bispectrum and the power spectrum. That we can extend the range of multipoles to less than $500$ means that we can use this formalism to make a prediction for $\tau_{NL}$ generated by cosmic strings in the multipole range of Planck.

\subsection{Estimator of $\tau_{NL}$}
The signal to noise ratio for the trispectrum (assuming the noise is cosmic variance dominated) is given by
\begin{eqnarray*}
(S/N)^2=\sum_{l_1\leq l_2 \leq l_3 \leq l_4}\sum_L \frac{|T_{l_1 l_2 l_3 l_4}(L)|^2}{(2L+1)C_{l_1}C_{l_2}C_{l_3}C_{l_4}}=\frac{1}{24} \sum_{l_1, l_2, l_3, l_4}\sum_L \frac{|T_{l_1 l_2 l_3 l_4}(L)|^2}{(2L+1)C_{l_1}C_{l_2}C_{l_3}C_{l_4}}
\end{eqnarray*}
where $T_{l_1 l_2 l_3 l_4}(L)$ is one of the possible configurations of the full-sky trispectrum. In order to find the signal to noise ratio we therefore must relate the full sky trispectrum to the flat sky equivlaent (\cite{Hu}).
\par
Decomposing the temperature perturbation in terms of spherical harmonics we have
\begin{eqnarray*}
\frac{\Delta T}{T}(\mathbf{\hat{n}})=a_{lm}Y_{lm}(\mathbf{\hat{n}}).
\end{eqnarray*}
In the flat sky limit the temperature perturbation is expressed in terms of Fourier harmonics
\begin{eqnarray*}
\frac{\Delta T}{T}(\mathbf{\hat{n}})=\int \frac{d^2 l}{(2\pi)^2} \delta(\mathbf{l_1})e^{i \mathbf{l.\hat{n}}}.
\end{eqnarray*}
The relationship between the full-sky and flat-sky limit is obtained by noting
\begin{eqnarray*}
a_{lm}&=&i^m \sqrt{\frac{2l+1}{4\pi}}\int \frac{d\phi_l}{2\pi}e^{i m \phi_l}\delta(\mathbf{l}),\\
\delta(\mathbf{l_1+l_2+l_3+l_4})&=&\int\frac{d\mathbf{\hat{n}}}{(2\pi)^2}	e^{i (\mathbf{l_1+l_2+l_3+l_4}).\mathbf{\hat{n}} }	\\
e^{i\mathbf{l,\hat{n}}}&=&\frac{2\pi}{l}\sum_m i^m Y_{lm}(\mathbf{\hat{n}})e^{i m \phi_l}\\
\left<a_{l_1 m_1}a_{l_2 m_2}a_{l_3 m_3}a_{l_4 m_4}\right>_c&=&\left(\Pi_{i=1}^4 \sqrt{\frac{l_i}{2\pi}}e^{-i m_i \phi_{l_i}}\right)(2\pi)^2 \delta(\mathbf{l_1+l_2+l_3+l_4}) T^{(l_1, l_2)}_{(l_3,l_4)}
\end{eqnarray*}
where $\phi_l$ denotes the polar angle of $\mathbf{l}$ and $T^{(l_1, l_2)}_{(l_3,l_4)}$ denotes the flat sky trispectrum given by a configuration with multipoles $l_1,l_2,l_3,l_4$ (and where $l_{12}$ is the diagonal of the quadrilateral that makes a triangle with $l_1,l_2$). Using these formulae we find
\begin{eqnarray*}
\delta((\mathbf{l_1+l_2+l_3+l_4}))&=&\int d^2 L \delta((\mathbf{l_1+l_2+L}))\delta((\mathbf{l_3+l_4-L}))\\
&=&\frac{1}{(2\pi)^2}\sum_{m_i}\sum_{L,M}\left(\Pi_{i=1}^4 \sqrt{\frac{2\pi}{l_i}}e^{i m_i \phi_{l_i}}	\right)\frac{2L+1}{4\pi}\times\\
&&\sqrt{(2l_1+1) (2l_2+1) (2l_3+1) (2l_4+1)}\left( \begin{array} {ccc}
l_1 & l_2 &L \\
0 & 0  &0\end{array} \right)\left( \begin{array} {ccc}
l_3 & l_4 &L \\
0 & 0  &0\end{array} \right) \times\\
&& (-1)^M\left( \begin{array} {ccc}
l_1 & l_2 &L \\
m_1& m_2  &M\end{array} \right)\left( \begin{array} {ccc}
l_3 & l_4 &L \\
m_3 & m_4  &-M\end{array} \right).
\end{eqnarray*}
Combining these results we can write the full sky trispectrum in terms of the flat sky trispectrum as
\begin{eqnarray}
T_{l_1 l_2 l_3 l_4}(L)=\frac{2L+1}{4\pi}\sqrt{(2l_1+1)(2l_2+1)(2l_3+1)(2l_4+1)}\left( \begin{array} {ccc}
l_1 & l_2 &L \\
0 & 0  &0\end{array} \right)\left( \begin{array} {ccc}
l_3 & l_4 &L \\
0 & 0  &0\end{array} \right) T^{(l_1 l_2)}_{(l_3 l_4)}(L).
\end{eqnarray}
With this consideration we can make an estimate of the signal to noise ratio for the trispectrum. We note also that we need only consider even terms since the Wigner-$3j$ vanishes for $l_i+l_j+L_{ij}= \mbox{odd}$. 
\par 
We are interested in finding an estimate for $\tau_{NL}$. We note that for the local model of inflation, \cite{Komatsu} (for which $\tau_{NL}=(6f_{NL}/5)^2$), the signal to noise ratio is given approximately by
\begin{eqnarray*}
(S/N)^{local}\approx 5\times 10^{-18} f_{NL}^{4 (local)} l_{max}^4 \approx 2.4\times 10^{-18}\tau_{NL}^{2(local)} l_{max}^4 
\end{eqnarray*}
Observationally the trispectrum has not received much attention with the rather
weak constraint $\tau_{NL} < 10^8$ \cite{whomever}.   Recently there has been
a significant improvement claimed using N-point probability distributions \cite{vielva}
with $-5.6\times 10^5<\tau_{NL}<6.4\times 10^5$.  Again these results are for a
local-type non-Gaussianity which is unlike the less peaked cosmic string trispectrum,
so a more specific analysis will be necessary to achieve quantitative constraints.

\section{Conclusions}

In this paper, we have endeavoured to analytically calculate the cosmic string
power spectrum, bispectrum and trispectrum over a range of scales relevant for
CMB experiments on both large and small angular scales.    We have been
particularly focused on extending previous work to multipole ranges ($l<2000$)
applicable for the Planck satellite,  an experiment which has the potential to
dramatically improve constraints on all these correlators.   We have presented
a relatively featureless shape for the bispectrum over the relevant range which should be easily
distinguishable from the oscillatory peaks of inflationary bispectra \cite{Fergusson} if there
is a significant signal discovered.    Our preliminary estimates of $f_{NL}$ from cosmic
strings indicate that Planck constraints from the bispectrum should be competitive with
those from the power spectrum.      We note that we obtain a considerably smaller estimate than
the $f_{NL} = -1000$ in ref.~\cite{Hindmarsh} for several reasons, including the extension
of  our analysis over lower multipole ranges relevant for WMAP, a more careful comparison
with  $f_{NL}$ estimators used in the literature, and a lower normalisation of the string
spectrum derived for Nambu strings \cite{Battye}, rather than that obtained from field
theory simulations \cite{Kunzetal}.   Nevertheless, as we shall discuss, our estimate
contains many uncertainties and more detailed and accurate forecasts are the subject
of ongoing work \cite{FergLandetal}.

We have also evaluated the CMB trispectrum (for
parallelogram configurations) on the relevant multipole scales for WMAP and Planck.
Again we find a relatively constant trispectrum, finite in all regimes and for which
we do not expect dramatic features for other configurations.    Our results indicate a
relatively larger signature  for the trispectrum,  in
contrast to the amplitude of the bispectrum which is suppressed because it depends on
the poor correlation between the string velocity and curvature.   Our preliminary estimate
of $\tau_{NL} $ a few times $10^4$ needs a more detailed analysis to characterise key
uncertainties more carefully.   Nevertheless, the trispectrum deserves much closer scrutiny observationally and the prospect of constraining cosmic strings should motivate the
development of suitable estimators.

These analytic calculations of the post-recombination gravitational effects of cosmic
strings offer important physical insights into the CMB correlations they induce.
 However, we note that there are many directions in which they
can be substantially improved.   Detailed numerical investigations of the CMB power spectrum
created by cosmic strings and other topological defects already takes into account
a far wider range of physical effects, notably recombination physics around decoupling
and a better description of the evolving string network.    It should be possible to
similarly develop the unequal time correlation methods presented for the late
time GKS signatures here to calculate both the bispectrum and the trispectrum to
high accuracy numerically.   In the meantime, however, it is a matter of comparing
the present analytic results with CMB maps induced by cosmic string networks on both
large (full sky) and small scales in order to get a more accurate normalisation and
characterisation of the bispectrum and trispectrum.   Given the stark contrast
between the string shapes with those predicted by inflation, there needs also to be
a specific search for these signatures in present and forthcoming CMB data,
a project which is being actively investigated \cite{FergLandetal}.   There seem
to be  good prospects of using forthcoming CMB data to obtain new insight into
cosmic string scenarios using higher order correlators.

\section{Acknowledgements}

We are very grateful for many informative discussions with 
James Fergusson and also for generating the three dimensional bispectrum images.  We also acknowledge useful
conversations with David Seery and Michele Liguori.
 EPS was supported by STFC grant ST/F002998/1 and the 
Centre for Theoretical Cosmology.  DMR was supported by EPSRC, the Isaac Newton Trust and the Cambridge European Trust.
We are grateful for the hospitality of the Tufts Taillores Centre during the Vilenkinfest in September 2009 where these results were first reported. 
\bigskip

\noindent {\it Note added in proof: While this manuscript was nearing completion, related results for the string trispectrum were presented in \cite{hind10}.}

\bibliographystyle{unsrt}
\bibliography{CosmicStringSpectra}

\end{document}